\documentclass[conference]{IEEEtran}
\IEEEoverridecommandlockouts

\usepackage{cite}
\usepackage{amsmath,amssymb,amsfonts}
\usepackage{algorithmic}
\usepackage{graphicx}
\usepackage{textcomp}
\usepackage{xcolor}

\usepackage{amssymb} 
\usepackage{cancel} 
\usepackage{pifont}
\usepackage{multirow}
\usepackage{booktabs}

\def\BibTeX{{\rm B\kern-.05em{\sc i\kern-.025em b}\kern-.08em
    T\kern-.1667em\lower.7ex\hbox{E}\kern-.125emX}}
\begin{document}

\title{SongTrans: An unified song transcription and \\ alignment method for lyrics and notes}

\author{\IEEEauthorblockN{Siwei Wu\textsuperscript{1,2,3,*}~\thanks{$^{*}$ Equal Authors}, Jinzheng He\textsuperscript{1,*},  
Ruibin Yuan\textsuperscript{3},  
Haojie Wei\textsuperscript{4}, 
Xipin Wei\textsuperscript{1},
Chenghua Lin\textsuperscript{2,3,$\dagger$}~\thanks{$^{\dagger}$ Equal Corresponding Authors},
Jin Xu\textsuperscript{1,$\dagger$},
Junyang Lin\textsuperscript{1}}
\IEEEauthorblockA{\textsuperscript{1}Alibaba Group \\ \textsuperscript{2}University of Manchester \\ \textsuperscript{3} Multimodal Art Projection Research Community \\ \textsuperscript{4} Renming University of China}
}

\maketitle

\begin{abstract}
The quantity of processed data is crucial for advancing the field of singing voice synthesis. While there are tools available for lyric or note transcription tasks, they all need pre-processed data which is relatively time-consuming (e.g., vocal and accompaniment separation).
Besides, most of these tools are designed to address a single task and struggle with aligning lyrics and notes (i.e.,  identifying the corresponding notes of each word in lyrics).
To address those challenges, we first design a pipeline by optimizing existing tools and annotating numerous lyric-note pairs of songs.
Then, based on the annotated data, we train a unified SongTrans model that can directly transcribe lyrics and notes while aligning them simultaneously, without requiring pre-processing songs. Our SongTrans model consists of two modules: (1) the \textbf{Autoregressive module} predicts the lyrics, along with the duration and note number corresponding to each word in a lyric. (2) the \textbf{Non-autoregressive module} predicts the pitch and duration of the notes.
Our experiments demonstrate that SongTrans achieves state-of-the-art (SOTA) results in both lyric and note transcription tasks. Furthermore, it is the first model capable of aligning lyrics with notes.
Experimental results demonstrate that the SongTrans model can effectively adapt to different types of songs (e.g., songs with accompaniment), showcasing its versatility for real-world applications.
\end{abstract}


\section{Introduction}
\label{sec:intro}

Music is an integral part of human life, which has led to an increasing interest among researchers in singing voice synthesis. Recent advancements in singing voice synthesis have been driven by Diffusion-based approaches~\cite{DBLP:conf/icassp/ZhangCXXZB22,DBLP:conf/interspeech/ZhangXL0GZG23,DBLP:conf/aaai/ZhangHLHXCDHZ24,DBLP:journals/corr/abs-2304-09116,DBLP:conf/acl/HeLYHCLZ23} and GAN-based networks~\cite{DBLP:journals/corr/abs-2009-01776,DBLP:conf/interspeech/ZhangZLL22}. Besides, there are many works based on LLMs~\cite{yuan2024chatmusician,deng2024composerx,zhuo2023lyricwhiz}. However, despite the availability of some open-source singing voice corpora~\cite{DBLP:conf/mm/HuangC0LCZ21,DBLP:conf/nips/ZhangLWDL0HHZCZ22,DBLP:conf/interspeech/WangWZWLXZXB22}, obtaining a large amount of annotated song data remains a significant challenge for researchers in this field.

On the one hand, current song annotation methods incur significant computational and time overheads, due to the fact that many approaches rely on voice separation to enhance their performance.
On the other hand, no tool can directly transcript and align both lyrics and notes for a song. 
While Whisper~\cite{DBLP:conf/icml/RadfordKXBMS23}, Qwen-Audio~\cite{bai2023qwen,chu2023qwen,chu2024qwen2} and Fastcorrect~\cite{leng2021fastcorrect} demonstrates robust multilingual speech transcription capabilities, but they are not specifically designed for the task of lyric transcription. 
ROSVOT~\cite{li2024robust} excels in note transcription but heavily depends on word duration predictions by Montreal Forced Aligner (MFA) and requires vocal inputs.
MFA also tends to mistakenly label unique vocal expressions (e.g., melisma) in a song as silence, which significantly impacts the accuracy of note recognition.
Previous works like VOCANO~\cite{DBLP:conf/ismir/Zih-SingS19,vocano} and MusicYOLO~\cite{DBLP:journals/taslp/WangTYXC23} can directly label the pitch and duration of notes, but their performance is relatively limited.

In addition to extensively gathering song and lyric pairs through internet, we design a pipeline to annotate and align note and lyric of a song. It combines word-level and phone-level filtering to eliminate noise from the lyric data and utilize the Ultimate Vocal Remover (UVR) ~\cite{takahashi2017multi} to extract vocals from songs. Then, we refine MFA to align the lyric and vocal by segmenting the vocal based on silent intervals, which can effectively distinguish the segments mistakenly identified as silence by MFA.
To streamline the cumbersome and complex process of lyric and note annotation, we use the data annotated by our pipeline to develop a SongTrans model, which can directly annotate lyrics and notes of a song, and align them without silence detection, or MFA.
The SongTrans employes a method that combines autoregressive and non-autoregressive approaches.
We find that the autoregressive approach has a strong capability to align vocals and lyrics, as well as vocals and notes, while the robustness of pitch prediction using non-autoregressive models is stronger.
Therefore, in the autoregressive module, we fine-tune two Whisper models to generate lyrics, as well as the corresponding word durations and note numbers.
As for the non-autoregressive module, we fine-tune the encoder of Whisper to predict the duration and pitch of each note, aligning them with the lyrics according to the predicted word durations and note numbers.

We evaluate the performance of our SongTrans model on the M4Singer dataset~\cite{DBLP:conf/nips/ZhangLWDL0HHZCZ22}. Compared to the existing lyric transcription models, SongTrans achieves SOTA performance. As for Note Transcription task, SongTrans model also demonstrates competitive results compared to previous methods. 
Notably, our model is the first to have the ability to align notes and lyrics with strong performance in accuracy.
We further conducted experiments by combining the training data from M4Singer with our labeled data in various proportions. Experimental results show that data labeled by our pipeline enhances the model's overall capability. Additionally, we demonstrate that our model can effectively label data under diverse settings, including raw songs, vocals of songs, and vocals segmented by silence.

\section{Data Processing Pipeline}
\label{sec:Pipeline}
We collect $58,144$ songs with lyrics and sentence-level timestamps, resulting in a dataset of $807,960$ sentence-level song-lyric pairs.

\subsection{Accompaniment Separation.}
To enhance the performance of our pipeline, we conduct UVR on 8 A100 GPUs for about 4 days to separate the accompaniment for $807,960$ samples. Subsequently, we utilize the isolated vocal for further annotation tasks. 

\subsection{Lyric Transcription.}
\label{lyric model}
There is a significant amount of noise in the lyrics, such as the inclusion of singer information at the beginning or the omission of an entire line of lyrics. Additionally, many users are not professional song creators, and the audio quality of the songs they upload is often poor.

To tackle these issues, we use keywords from the gathered data to select songs by mainstream artists. We then employ the Whisper model to transcribe the lyrics of the song and regard it as whisper-trans lyrics.
Following this, we calculate the Word Error Rate (WER) between the original crawled lyrics and the whisper-trans lyrics. Additionally, we translate the lyrics into phones and compute the WER at the phone level.

\textbf{Word-Level Filtering.} If the word-level WER is less than 0.3, we consider the lyrics of the samples obtained by web crawling to be accurate.

\textbf{Phone-Level Filtering.} The whisper-trans lyrics may differ from the correct lyrics, but their pronunciation remains very similar.
Therefore, we also accept the samples whose word-level WER is higher than 0.3 but with a phone-level WER less than 0.4.

We randomly select 50 samples and manually evaluate our method of different thresholds on it. The thresholds (i.e., 0.3 at word level and 0.4 at phone level) are the maximum values that can achieve an accuracy of 80\%.
Finally, we obtain $201,649$ sentence-level song-lyrics pairs and train a lyric-trans model by fine-tuning Whisper model.

\subsection{MFA Refinement.}
To obtain the duration of each word, we use the MFA to align the word with the song.
However, as for the singing data, there are some special vocal techniques such as melisma, where MFA tends to recognize melisma as silence.
On the one hand, the alignment between lyrics and notes highly relies on the result of MFA. On the other hand, the mistakenly recognized melisma portion will be confused with the existing silence in the audio and cannot be identified.

Therefore, we re-segment the song by detecting the silent part within it.
Then we use the lyric-trans model to relabel the resegmented song and obtain $273,282$ lyric-song pairs.
We use the MFA to align the latest lyric-song pairs and merge the silence part predicted by the MFA with the preceding phone.

\subsection{Note Labelling.}
We use ROSVOT~\cite{li2024robust}, a SOTA model in note transcription task, to label the notes of a song.
However, ROSVOT sometimes predicts the phone at the beginning or the end of a sentence as silence (i.e., pitch is 0).
We find most pitches of adjacent phones are close to each other, so we directly use the pitch of the neighboring phone as the pitch of the phone which is mistakenly recognized as silence. 

\begin{figure*}[!tb]
    \centering
    \includegraphics[width=13cm]{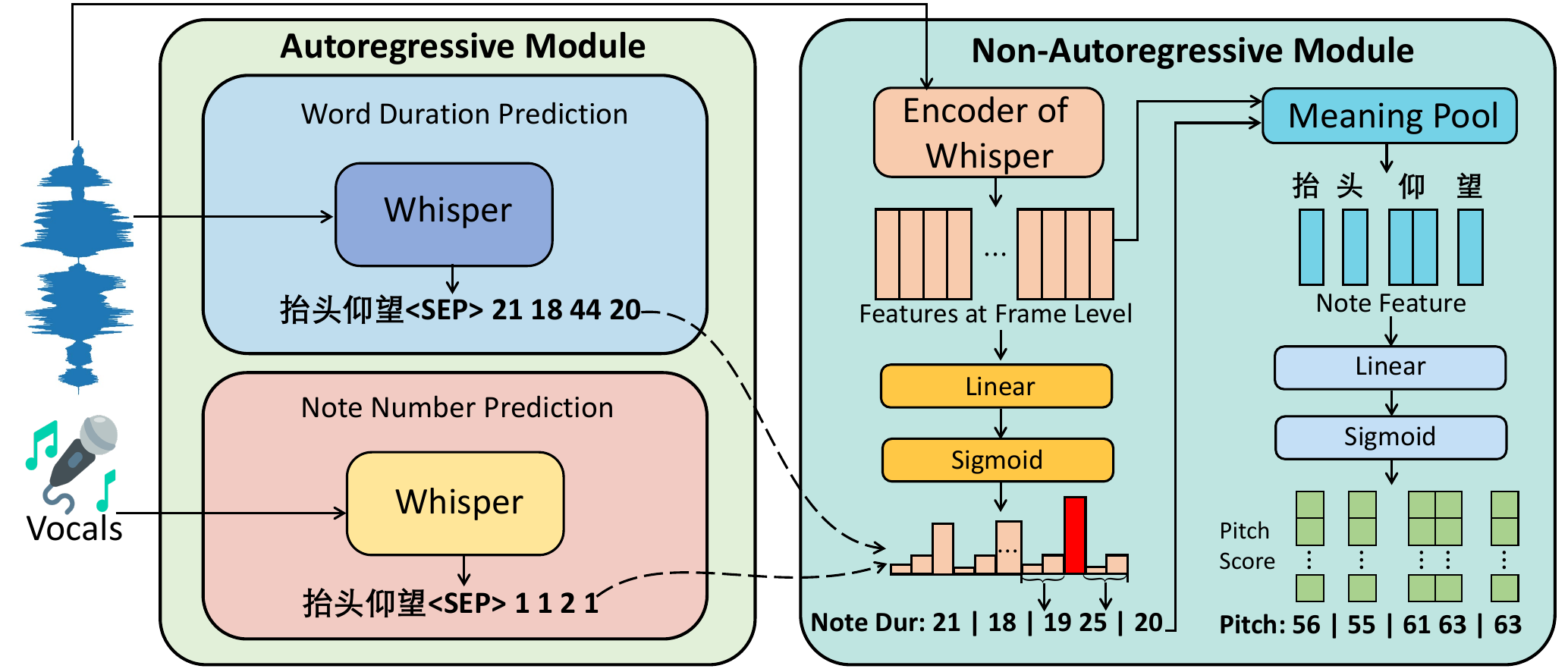}
    \caption{The framework of our SongTrans Model.}
    \label{fig:model framework}
\end{figure*}

\section{Method}
\label{sec:pagestyle}

Using the lyric-note-song triples annotated by our pipeline, we train a SongTrans model to automatically label and align the lyrics and notes. 
Our SongTrans model consists of two components: the Autoregressive Module and the Non-Autoregressive Module, as illustrated in Fig~\ref{fig:model framework}. 

\subsection{Autoregressive Module}
\label{sec: Autoregressive Module}
In this module, we fine-tune two Whisper models to end-to-end generate lyric and lyric duration pairs, as well as lyric and note number pairs.

\noindent\textbf{(1) Lyric and Lyric Duration Transcription. }
We fine-tune Whisper to generate lyric and word duration end-to-end. 
Specifically, we design the output of the model in the format of `[Lyric]\texttt{<}SEP\texttt{>}[Word\_Duration]', where the \texttt{<}SEP\texttt{>} is the special token, [Lyric] presents the sequence of lyric, and [Word\_Duration] is the sequence to present the duration of each word with the same length of [Lyric].

\noindent\textbf{(2) Note Number Detection.} 
Singing is different from the speech due to each syllable may correspond to multiple notes.
We train another Whisper model to generate a sequence of note numbers corresponding to the lyrics.
We define the output of the model in the format of `[Lyric]\texttt{<}SEP\texttt{>}[Note\_Number]', and the [Note\_Number] is a sequence presents the number of notes for each word in a lyric.

We fine-tune Whisper by cross-entropy loss of the next word.

\subsection{Non-Autoregressive Module}
In this part, we fine-tune the Encoder of a Whisper to predict the duration and the pitch of the note.

We input the audio of a song $X$ into the encoder of the Whisper to obtain the feature of the fragments:

\begin{equation}
    [\textbf{v}_0, \textbf{v}_1, ..., \textbf{v}_n] = \mathrm{WhisperEcoder}(X),
\end{equation}
where $n$ represent the max length of the audio and $v_i$ is the feature of i-th fragment $f_i$.

As for the note duration prediction, we use Whisper's encoder to predict the boundary of each note within the duration of a word. We input the fragment feature $v_i$, and use linear and sigmoid to calculate the boundary probability:

\begin{equation}
    p_{i} = \mathrm{Sigmoid}(\mathrm{Linear}(\textbf{v}_i)),
\end{equation}
where $p_i$ represents the probability that fragment $i$ is the boundary of a note.
Then, we use the Binary Cross Entropy (BCE) to calculate the loss for the boundary prediction:

\begin{equation}
    Loss_b = \mathrm{BCE(\textbf{p}, \textbf{b}_{gt})},
\end{equation}
where $\textbf{p}$ represent the probability of all fragments and $\textbf{b}_{gt}$ is the ground truth note boundary of a song.
During inference, we use the note number $k$ and word duration predicted by Sec~\ref{sec: Autoregressive Module} to select the top $k-1$ fragments with the highest note boundary probabilities within the corresponding word duration.

As for the note pitch prediction, based on the predicted note boundary, we average the fragment features of a note to obtain the note feature $\textbf{v}_{mean}$. 
We then use a linear and sigmoid function to predict the pitch of a note:
\begin{equation}
    \textbf{p}_{note} = \mathrm{Sigmoid}(\mathrm{Linear}(\textbf{v}_{mean})),
\end{equation}
where $p_{note\_{i}}$ represents the probability that the note's pitch is $pitch_{i}$.
Similarly, we use the BCE to calculate the note prediction loss:

\begin{equation}
    Loss_{n} = \mathrm{BCE(\textbf{p}_{note}, \textbf{pitch}_{gt})},
\end{equation}
where the $\mathrm{\textbf{pitch}_{gt}}$ is the ground truth pitch label of note.

\begin{table*}[hbt!]
\begin{center} 
\footnotesize
\resizebox{1.9\columnwidth}{!}{
    \begin{tabular}{l|cc|cc|cc}
\toprule
\multicolumn{1}{l|}{ }                                                                     & \multicolumn{2}{c|}{\textbf{Lyric Trans (WER)}  }                                                                              & \multicolumn{2}{c|}{\textbf{Alignment (MAE)}}                                                                           & \multicolumn{2}{c}{\textbf{Note Trans}}                                                                            \\
\multirow{-2}{*}{\textbf{Model}}                                                                      & Lyric (\%) & Ph (\%) & Lyric Dur (0.01 s) & 
Note Num (0.01 cot) & Pitch of Note WER (\%) & Note Dur MAE (0.01 s) \\

\midrule\midrule
   Whisper  & 32.56 & 23.98	& \textcolor{red}{\ding{56}}  &	\textcolor{red}{\ding{56}} & \textcolor{red}{\ding{56}} & \textcolor{red}{\ding{56}}
\\
   MFA  & \textcolor{red}{\ding{56}} & \textcolor{red}{\ding{56}}	& 14.74  &	\textcolor{red}{\ding{56}} & 	 \textcolor{red}{\ding{56}} & \textcolor{red}{\ding{56}}
\\
    VOCANO  & \textcolor{red}{\ding{56}} & 	\textcolor{red}{\ding{56}}	& \textcolor{red}{\ding{56}} &  \textcolor{red}{\ding{56}} &	45.75 &  21.26
\\
   ROSVOT & \textcolor{red}{\ding{56}} & \textcolor{red}{\ding{56}} & \textcolor{red}{\ding{56}} & \textcolor{red}{\ding{56}} &  22.60\textsuperscript{*} & \textcolor{red}{\ding{56}} 
\\
\midrule
    Ours  & \textbf{10.81 }  & \textbf{	9.14}	& \textbf{10.80} &	\textbf{14.38} & \textbf{22.45} & \textbf{16.17}
\\
\bottomrule
\end{tabular}
}
\end{center}
\caption{
Overall results. The \textcolor{red}{\ding{56}} indicates that model does not have the corresponding functionality. The \textsuperscript{*} means that we report the results are directly copied from the original paper.
The unit of Note Num, 0.01 cot, represents 0.01 count of note number.
}
\label{tab:Main Result}
\end{table*}

\section{Experiments}

\subsection{Dataset and Backbone}
We separately designated two 3\% subsets of the M4Singer dataset as the test set and validation set for evaluating our model's performance. However, our labeled data consists of songs without the silence parts. To enable our model to adapt to the formal vocal audio containing silent parts, we propose a \textbf{data merging} method. This method combines our labeled data with the remaining M4Singer data for joint training.
Besides, the version of Whisper used in our work is Whisper-v3-1.5B.

\subsection{Baselines}
We compare our SongTrans model with the following models on the corresponding task: 
(1) \textbf{Whisper.} It is an autoregressive model for lyric transcription tasks.
(2) \textbf{ROSVOT.} ROSVOT is an AST model that serves Singing Voice Synthesis (SVS) by labeling the notes of songs.
(3) \textbf{VOCANO~\cite{vocano}.} It is a note transcription framework for singing voice in polyphonic music.
(4) \textbf{Montreal Forced Align (MFA).} It is a technique to take an orthographic transcription of an audio file and generate a time-aligned version using a pronunciation dictionary to look up phones for words.

Because the ROSVOT is trained on M4Singer and does not release their test split, we report results in original paper.

\subsection{Metric}
We mainly use the Mean Absolute Error (MAE) and Word Error Rate (WER) to evaluate the capability of the models in our compared system on each task.
\textbf{As for Dur MAE}, we align the corresponding predicted and ground-truth duration sequence by padding (i.e., 0) at the end.

\section{Results}
\subsection{Overall Performance Analysis}
As demonstrated in Table~\ref{tab:Main Result}, our SongTrans model is the first to achieve comprehensive functionality for both transcription and alignment of lyrics and notes.
As for the Lyric Transcription task, our method sets a new SOTA standard, drastically reducing WER by more than half compared to Whisper. Specifically, our model achieves WER of 9.14\% at the phone level and 10.80\% at the character level, in contrast to Whisper's 23.98\% and 32.56\%, respectively.
Moreover, at the note level, our performance is on par with the former note transcription methods, ROSVOT and VOCANO. Notably, our SongTrans is the first model to realize the ability to align both notes and lyrics. 
Our approach involves predicting the number of notes for each word in the lyrics, with a WER of just 0.108 s and an MAE of that is 0.1438 number prediction error per word.

\subsection{Impact of Data Format}

\begin{table}[!htp]
    \centering
    \resizebox{0.99\columnwidth}{!}{
        \begin{tabular}{cccc}
            \hline
            Setting & Lyric (WER) & Pitch of Note (WER) & Note Dur (MAE) \\ \hline
            w/o Seg & 10.81 & 22.45 & 16.17 \\ 
            w/ Seg & \textbf{8.61} & \textbf{21.32} & \textbf{11.62} \\ \hline
        \end{tabular}
    }
    \caption{The result of our SongTrans model on vocal data with and without the silence segmentation (Seg).}
    \label{tab: seg setting}
\end{table}

Because our data labeling pipeline segments song vocals by silence and the SongTrans model is trained on it, the presence of both silence and accompaniment significantly impacts our model's performance. Consequently, we also evaluate our model in two settings: (1) conducting silence segmentation and (2) adding accompaniment to a song.

As shown in Table~\ref{tab: seg setting}, we compare the performance of our SongTrans model with and without silence segmentation. Our model achieves SOTA performance when silence segments are removed. Notably, its performance on raw vocal audio, without silence segment processing, remains very close to the results on the w/ Seg setting. This indicates that our model is robust and capable of accurately recognizing both lyrics and notes in vocal audio without silence segmentation.

\begin{table}[!htp]
    \centering
    \resizebox{0.9\columnwidth}{!}{
        \begin{tabular}{ccc}
            \hline
            Ph (WER)  &  Pitch of Note (MAE)  & Note Dur (MAE)  \\ \hline
             14.01 &  2.69 & 25.56 \\ \hline
        \end{tabular}
    }
    \caption{Our model's result on songs with accompany.}
    \label{tab: accomany result}
\end{table}

Additionally, we demonstrate that our model can directly label both lyrics and notes for songs with accompaniment.
We perform vocal-accompaniment separation and apply our model to label lyrics and notes on a test set of our web-crawled data, which contains 50 songs.
We regard the notes and lyrics annotated by our model under the accompaniment separation setting as pseudo-golden answers.
Subsequently, we utilized our SongTrans model to predict lyrics and notes on the test set with accompaniment and compared the results with the pseudo-golden answers.
As shown in Table~\ref{tab: accomany result}, across various tasks, the results predicted by SongTrans with accompaniment are very close to the pseudo-golden answers.

All of the above observations demonstrate that the SongTrans model is capable of predicting lyrics and notes on various types of data, i.e., with accompaniment or without silence segmentation.

\subsection{Impact of Data Merge}

In real-world scenarios, songs are not segmented to exclude silence, which is different from the data annotated by our pipeline.
To adapt our model for real-world scenarios, we merged our labeled data with the training data from M4singer, which includes the silent parts. Then we train the SongTrans model on this combined dataset.
We also design experiments to demonstrate the effectiveness of our \textbf{data merge} method.

\noindent\textbf{(1) M4Singer data effectively enhances SongTrans' ability to handle data with silence.}
As illustrated in Table~\ref{tab: impact of M4Singer}, the incorporation of the M4Singer data leads to a significant enhancement in the performance of the SongTrans model on song data featuring silent parts. This result effectively validates our data merge strategy.

\begin{table}[t]
    \centering
    \resizebox{0.99\columnwidth}{!}{
        \begin{tabular}{cccc}
            \hline
            Training Data & Lyric (WER) & Pitch of Note (WER) & Note Dur (MAE) \\ \hline
            Ours & 23.44 & 34.64 & 24.39 \\ 
            Ours + M4Singer & \textbf{10.81} & \textbf{22.45} & \textbf{16.17} \\ \hline
        \end{tabular}
    }
    \caption{The performance of SongTrans model training with and without the M4Singer data. The Ours represents the data labeled by our pipeline.}
    \label{tab: impact of M4Singer}
\end{table}

\begin{figure}[tp]
\centering
\includegraphics[width=0.85\columnwidth]{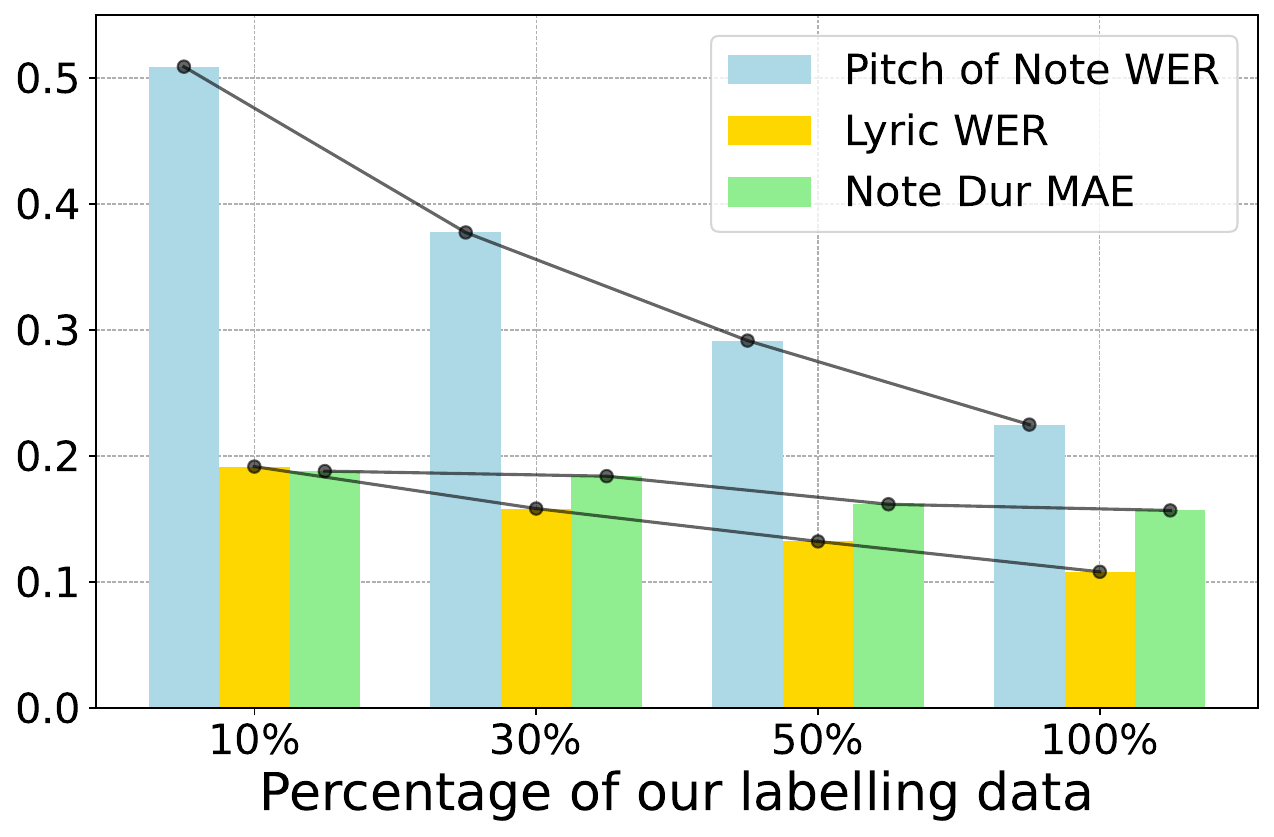}
\caption{The performance of SongTrans trained by merging varying percentages of our labeling data with M4Singer.}
\label{figure: merge data}
\end{figure}

\noindent\textbf{(2) Our labeled data can effectively enhance the performance of the SongTrans model.} We conducted experiments to train the model by merging various proportions of our labeled data with M4singer's training data. 
As shown in Fig~\ref{figure: merge data}, with the proportion of our labeled data increasing, the Pitch of Note WER, Lyric WER and Note Dur MAE decreases accordingly.
This shows the effectiveness of our labeling data. 

\end{document}